\documentclass[prd,twocolumn,showpacs,amsmath,amssymb]{revtex4}
\usepackage{graphicx}
\usepackage{xcolor}
\usepackage{dcolumn}
\usepackage{bm}
\newcommand{\EE}{e^+e^-}
\newcommand{\psp}{\psi^{\prime}}

\newcommand{\jpsi}{J/\psi}

\newcommand{\Pp}{P^{\prime}}
\newcommand{\ag}{a_{3g}}
\newcommand{\aga}{a_{\gamma}}
\newcommand{\ac}{a_c}

\newcommand{\pip}{\pi^+}
\newcommand{\pim}{\pi^-}
\newcommand{\piz}{\pi^0}
\newcommand{\rop}{\rho^+}
\newcommand{\rom}{\rho^-}
\newcommand{\roz}{\rho^0}
\newcommand{\kap}{K^+}
\newcommand{\kam}{K^-}
\newcommand{\kaz}{K^0}
\newcommand{\kazb}{\overline{K}^0}

\newcommand{\kstp}{K^{*+}}
\newcommand{\kstm}{K^{*-}}
\newcommand{\kstz}{K^{*0}}
\newcommand{\kstzb}{\overline{K}^{*0}}
\newcommand{\etap}{\eta^{\prime}}
\newcommand{\ampM}{{\cal M}}
\newcommand{\ffkt}{{\cal F}}
\newcommand{\Heff}{{\cal H}_{eff}}

\newcommand{\gz}{g_{0}}
\newcommand{\beq}{\begin{equation}}
\newcommand{\eeq}{\end{equation}}
\def\eref#1{(\ref{#1})}
\def\Journal#1#2#3#4{{#1} {\bf #2}, #3 (#4)}
\def\PLB{Phys. Lett. B}
\def\PRD{Phys. Rev. D}

\begin{document}

\title{Cross section and parametrization of charmonium decay}

\author{Xiao-Hu MO$^{1,3,4}$, Jin-Tao CHEN$^{2}$, You-Kai WANG$^{2}$
\\  \vspace{0.2cm} {\it
$^{1}$ Institute of High Energy Physics, CAS, Beijing 100049, China\\
$^{2}$ School of Physics and Information Technology, Shanxi Normal University, Xi'an 710119, China\\
$^{3}$ University of Chinese Academy of Sciences, Beijing 100049, China\\
$^{4}$ State Key Laboratory of Particle Detection and Electronics, Beijing 100049, Hefei 230026, China\\}
}
\email{moxh@ihep.ac.cn}
\date{\today}
	
\begin{abstract}
The parametrization forms of charmonium quasi two body decays are discussed in detail in this paper. The symmetry analysis and magnetic transition description are utilized to provide the multi-aspect comprehension of the decay dynamics, including the meson mixing angle, form factor, and $SU(3)$ symmetry breading effect. As the prerequisite, the electromagnetic cross sections involving scalar and pseudoscalar, are calculated for eight final states based on the approach of current algebra. Moreover, the mathematical manipulations of calculation for four kinds of final states are spelled out to illuminate the technique strategy.
\end{abstract}
\maketitle

\section{Introduction}
More and more charm and charmonium data in the world have been collected since the upgraded Beijing Electron-Positron Collider (BEPCII) and spectrometer detector (BESIII) started data taking in 2008~\cite{bes,yellow}. The high-statistics data samples, especially the data at $\jpsi$ and $\psp$ resonance peaks, which provide an unprecedented opportunity to acquire useful information for understanding the interaction dynamics by analyzing various kinds of decay final states.

It is well known that quantum chromodynamics (QCD) as a widely appreciated theory of strong interaction, has been proved to be very successful at high energy when the calculation can be executed perturbatively. Nevertheless, its usage at the nonperturbative regime, such as $\jpsi$ and $\psp$ resonance regions, needs more experimental guidance. At present, the phenomenological analysis plays a crucial role for enriching our understanding of the general principle. The essence of such a analysis is to make use of group theory as much as possible in order to exhaust the symmetry properties and leave our ignorance of dynamical details with undetermined parameters and energy-dependent functions.

As a matter of fact, many phenomenological models are constructed for charmonium decay~\cite{Kowalski:1976mc}-\cite{moxh2025}, the parametrization forms of various decay modes are obtained, such as the pseudoscalar and pseudoscalar mesons (PP), vector and pseudoscalar mesons (VP), octet baryon-pair, and so on. Recently, a systematical parametrization scheme has been proposed~\cite{moxh2023,moxh2024,moxh2025}, by virtue of $SU(3)$ flavor symmetry, the effective interaction Hamiltonian in tensor form is obtained according to group representation theory. In the light of flavor singlet principle, a uniform approach of parametrization is realized for all charmonium decays, involving binary decays, ternary decays and radiative decays. The parameters appeared in the parametrization scheme, which can be figured out by virtue of data analysis, will shed the light on the dynamics of charmonium decay.

However, for data analysis the collision cross section is the key tie between the phenomenological model and the experimental information. Moreover, the differential cross section that contains the angular distribution of particles usually serves as a signature for the spin of a particle. Generally speaking, there are three approaches to calculate the cross sections, the first is helicity formalism, one commonly used form of which has been established by Suh-Urk Chung in a series of papers~\cite{chungsu2008,chungsu1998,chungsu1993,chungsu1971}. The general validity of this formalism lies in that it can be adopted for any mass and spin of mesonic particles involved in a decay. A general scheme of $lS$-coupling analysis is developed to construct the Lorentz invariant amplitude. The clear physical and geometric picture make such a formalism is extensively adopted in partial wave analysis.

The second approach is tensor formalism~\cite{filippini1995,zou2003}, the obvious merit of which is that it is fully covariant and can be applied easily to many body decays. Another merit is that it is easier to code and timesaving for actual program computation.

The third approach is current formalism, by virtue of which we will calculate eight kinds of cross sections of
of charmonium decay~\cite{tosa1976}. In the next section, the details of current formalism is expound, and four types of cross sections are used to exemplify the evaluation procedure. The section that follows will focus on the parametrization of VP final state. The general symmetry analysis scheme is discussed for the description of coupling constants, and the radiative transition amplitude is used to figure out the structural feature of constitute quarks. After that is a summary section. Relegated to the appendix are materials relevant to various calculation details.

\section{Cross section}\label{xct_krxns}
In this paper, we only consider the two body final state, that is, the annihilation processes of $e^+$ and $e^-$ take place through the conversion of the pair into a virtual photon with the mass equal to the center-of-mass system (CMS) energy, then the photon converts into the two mesons final state. The total differential cross section reads~\cite{quangpham}
\beq
\frac{d\sigma (e^+ + e^-  \to M_1+M_2)}{d\Omega} = \frac{|{\cal M}|^2}{64 \pi^2 s } \cdot \frac{|\vec{k}|}{|\vec{p}|}~,
\label{ttdffxn}
\eeq
where $\sqrt{s}$ is the CMS energy and $\vec{k}$ ($\vec{p}$) is the momentum of either the $M_1$ ($e^+$) or the $M_2$ ($e^-$), and the transition amplitude ${\cal M}$ is shorthand for $\langle f | \hat{\cal M} | i \rangle$. Here $| i \rangle (=| e^+ e^- \rangle)$ and $| f \rangle (=| M_1 M_2 \rangle)$ indicate the initial and final states, respectively; the operator $\hat{\cal M}$ encapsulates all dynamic content of the interaction. Since all quantities of $|{\cal M}|^2$ are to be expressed in the CMS, some commonly used quantities are summarized below.

First, the four kinematic variables are
\beq
\begin{array}{c}
p_1 =(E,E\hat{z})~,~~ p_2 =(E,-E\hat{z})~, \\
k_1 =(E_1,\vec{k})~,~~ k_2 =(E_2,-\vec{k})~,
\end{array}
\label{ffmmtum}
\eeq
where $p_1$ and $p_2$ ($k_1$ and $k_2$) are four-momentum vectors of $e^+$ and $e^-$ ($M_1$ and $M_2$), respectively. $\hat{z}$ indicates the unit vector of $z$-axis, which is also the beam direction.
Hereinafter, the mass of electron or positron ($m_e$) is commonly neglected, and $E=|\vec{p}|$ is always true in this paper. By virtue of Eq.~\eref{ffmmtum}, it yields at once that
\beq
\begin{array}{rcl}
p_1 \cdot k_1 &=& E E_1 - E |\vec{k}| \cos \theta ~, \\
p_2 \cdot k_1 &=& E E_1 + E |\vec{k}| \cos \theta ~, \\
p_1 \cdot k_2 &=& E E_2 + E |\vec{k}| \cos \theta ~, \\
p_2 \cdot k_2 &=& E E_2 - E |\vec{k}| \cos \theta ~, \\
k_1 \cdot k_1 &=& E_1^2 - |\vec{k}|^2 =m_1^2 ~, \\
k_2 \cdot k_2 &=& E_2^2 - |\vec{k}|^2 =m_2^2 ~, \\
k_1 \cdot k_2 &=& E_1 E_2 + |\vec{k}|^2 ~, \\
p_1 \cdot p_2 &=& 2 E^2~,
\end{array}
\label{pkdotxpna}
\eeq
where $\theta$ is angle between $\hat{z}$ and $\vec{k}$ (i.e. $\hat{z} \cdot \vec{k}= |\vec{k}| \cos \theta$).
If we define
\beq
q \equiv p_1+p_2~,~~ k \equiv k_1+k_2~,
\label{defqandk}
\eeq
the conservation of energy and momentum leads to
\beq
q = k = (2E,\vec{0})~,
\label{qandkrln}
\eeq
then it yields
\beq
\begin{array}{c}
k \cdot p_1 = 2E^2~,~~ k \cdot k_1 = 2E E_1~, \\
k \cdot p_2 = 2E^2~,~~ k \cdot k_2 = 2E E_2~. \\
\end{array}
\label{kpdotxpnb}
\eeq

The following work is devoted to the calculation of $|{\cal M}|^2$. We will firstly construct the covariant vertex of certain decay, from which the current is extracted. Secondly, it is a matter of algebraic manipulations about the contraction for various four-vectors. At last, taking advantage of the special quantities in CMS, the cross section is expressed by experimentally measurable variables. Some subtleties of the details have to be considered carefully so that the differential cross section can be acquired accurately. The four kinds of final states, that is pseudoscalar-pseudoscalar (PP), pseudoscalar-vector (PV), pseudoscalar-axial-vector (PA), and pseudoscalar-tensor (PT) final states, will be used to display the technological process of calculation.

\subsection{PP final state}\label{xct_xnppfs}
\begin{figure}[htb]
\includegraphics[width=7.cm]{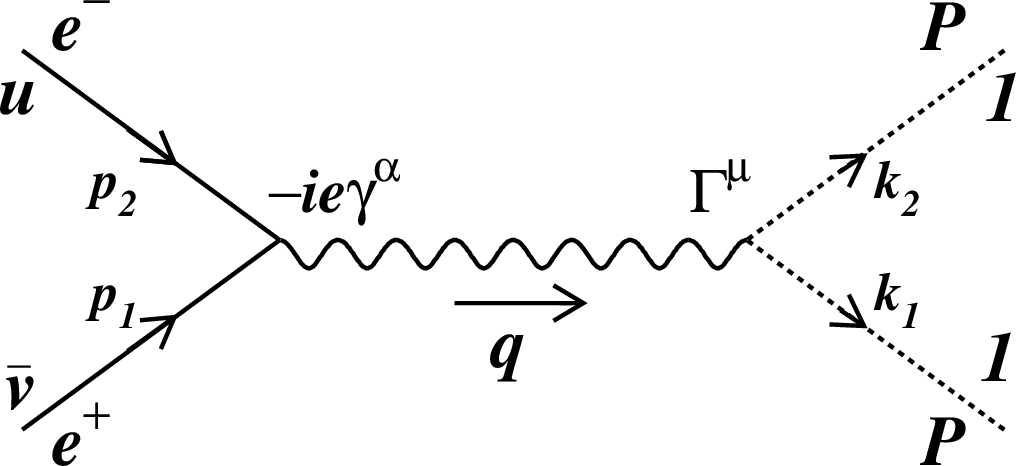}
\caption{\label{fig_ppfs}Feynman diagram of $\EE  \to PP$ process.}
\end{figure}

The Feynman diagram of $\EE \to PP$ process is shown in Fig.~\ref{fig_ppfs}. We begin with constructing the covariant vertex~\cite{tosa1976}, which has the form
\beq
F_{\mu \nu } \partial^{\mu} P \partial^{\nu} \Pp ~.
\label{vtxppfs}
\eeq
Expressing the vertex in momentum representation, we have
\beq
\begin{array}{rcl}
F_{\mu \nu } \partial^{\mu} P \partial^{\nu} \Pp
    &=& (\partial_{\mu} U_{\nu} - \partial_{\nu} U_{\mu} ) \partial^{\mu} P \partial^{\nu} \Pp \\
    &=& i(k_{\mu} U_{\nu}-k_{\nu} U_{\mu}) k_1^{\mu}k_2^{\nu} P_1 P_2 \\
    &=& i [(k \cdot k_1) k_{2 \mu} - (k \cdot k_2) k_{1 \mu} ] U^{\mu} P_1 P_2~,
\end{array}
\label{vtxppxpn}
\eeq
from which the current is obtained
\beq
\Gamma^{\gamma P P}_{\mu}= i~ e~ [(k \cdot k_1 ) k_{2\mu}-(k \cdot k_2) k_{1\mu} ]~.
\label{krtppfs}
\eeq

Some remarks are in order here. First, the above current is solely valid for pointlike particles. So far as a meson is concerned, it is composed of quarks, which has inter feature and construction. The electromagnetic interaction of a meson with a virtue photon is conventionally described by a form factor $f_{PP}(q^2)$. This factor should be added in formula~\eref{krtppfs}. However, it is a fix factor in the calculation of the amplitude, therefore the explicit reference to such a factor is dropped in this section and will be restored when we discuss the parametrization of charmonium decay. Second, when we construct the covariant vertex, sometimes the extra momentum operators have to be introduced. Nevertheless, since the $|{\cal M}^{tot}|^2$ is essentially a dimensionless quantity, extra dimensions must be normalized out. Therefore, a normalization factor is introduced as follows
\beq
\xi = \frac{1}{E^{2n}}~,
\label{nmfactor}
\eeq
where $E=\sqrt{s}/2$ is the beam energy and $n$ is the number of momentum operators. For PP final state, $n$ equals to 2. This factor should be added in formula~\eref{krtppfs} as well. However, for the concise of calculation, it is taken into account in the final step when the differential cross section is evaluated.

By making use of Feynman rules and referring to Fig.~\ref{fig_ppfs}, the covariant amplitude is cast as
\beq
\begin{array}{rcl}
i\ampM&=& {\displaystyle \bar{v}(-ie\gamma^{\rho}) u \cdot \frac{-ig_{\rho \mu}}{q^2+i\epsilon}
       \cdot \Gamma^{\gamma P P}_{\mu} P \Pp }\\
    &=&{\displaystyle - i\frac{e^2}{q^2}(\bar{v}\gamma^{\mu} u)[(k \cdot k_1 ) k_{2\mu}-(k \cdot k_2) k_{1\mu}]~. }
\end{array}
\label{ampofpp}
\eeq
For unpolarized beams and undetected final spins, one sums over all final spin states and averages over all initial spin states. This spin averaging will introduce a factor of $1/4$, since the electron and the positron have two possible spin states each. Herein, we introduce the notation

\beq
|\overline {\ampM }|^2= \frac{1}{4} \sum^{inital}_{all~spins} \sum^{final}_{all~spins} |i \ampM |^2~.
\label{ttampofpp}
\eeq
Exploiting Casimir's trick~\cite{griffiths},
\beq
\begin{array}{rcl}
\sum\limits_{all~spins} [\bar{v}\gamma^{\mu} u\bar{u}\gamma^{\nu}v] &=&
\mbox{Tr} [\gamma^{\mu} (\rlap/p_1 +m_e) \gamma^{\nu} (\rlap/p_2 +m_e) ] \\
   &=& 4 [ p^{\mu}_1 p^{\nu}_2+p^{\nu}_1 p^{\mu}_2-g^{\mu\nu} (p_1 \cdot p_2)]~.
\end{array}
\label{traceofee}
\eeq
At the last step, the term relevant to $m^2_e$ has been neglected as we claimed before. Then
\beq
\begin{array}{rcl}
|\overline {\ampM }|^2 &=&{\displaystyle \frac{e^4}{q^4} [ p^{\mu}_1 p^{\nu}_2+p^{\nu}_1 p^{\mu}_2-g^{\mu\nu} (p_1 \cdot p_2)] } \\
 & \times & [(k \cdot k_1)^2 \cdot k_{2 \mu} k_{2 \nu} +(k\cdot k_2)^2\cdot k_{1 \mu} k_{1 \nu} \\
 & & -(k\cdot k_1)(k \cdot k_2)(k_{1 \mu} k_{2 \nu} +k_{1 \nu} k_{2 \mu})] \\
 &=&{\displaystyle \frac{e^4}{q^4} \{ (k \cdot k_1)^2 [2~ k_2 \cdot p_2~ k_2 \cdot p_1 + k_2^2~ p_1 \cdot p_2 ] } \\
 & &+(k \cdot k_2)^2 [2~ k_1 \cdot p_2~ k_1 \cdot p_1 + k_1^2~ p_1 \cdot p_2 ] \\
 & &-2~ k\cdot k_1 ~ k \cdot k_2 ~ [ k_1 \cdot p_1 ~k_2 \cdot p_2  \\
 & & + k_1 \cdot p_2 ~k_2 \cdot p_1 - k_1 \cdot k_2 ~p_1 \cdot p_2 ] \}
\end{array}
\label{ttampofppa}
\eeq
Substituting the above terms with the corresponding values in CMS as given in Eqs.~\eref{pkdotxpna} and \eref{kpdotxpnb}, it yields
\beq
|\overline {\ampM }|^2= 2 e^4 E^2 |\vec{k}|^2 (1-\cos^2 \theta)~.
\label{ttampofppb}
\eeq
Taking the normalization factor $\xi=1/E^4$ into consideration, we derive the differential cross section
\beq
\frac{d\sigma_{PP} }{d\Omega} = \frac{\alpha^2 \beta^3}{2 s } \cdot (1-\cos^2 \theta)~,
\label{dffxnpp}
\eeq
where
\beq
\alpha \equiv \frac{e^2}{4 \pi}~,~~\beta \equiv \frac{|\vec{k}|}{E} ~,
\label{defalfabta}
\eeq
with
\beq
|\vec{k}| = \frac{[(s-(m_1+m_2)^2)(s-(m_1-m_2)^2)]^{1/2}}{2\sqrt{s}}~.
\label{vectork}
\eeq
Here $m_1$ and $m_2$ are the masses of two mesons. When $m_1=m_2=m$,
\beq
\beta =\sqrt{1-\frac{4 m^2}{s}} ~,
\label{betavalue}
\eeq
which is just the velocity of meson. The integration over the angular variables leads to the total cross section
\beq
\sigma_{PP} (s) = \frac{4 \pi \alpha^2 \beta^3}{3 s }~.
\label{ttxnpp}
\eeq

\subsection{PV final state}
\begin{figure}[htb]
\includegraphics[width=7.cm]{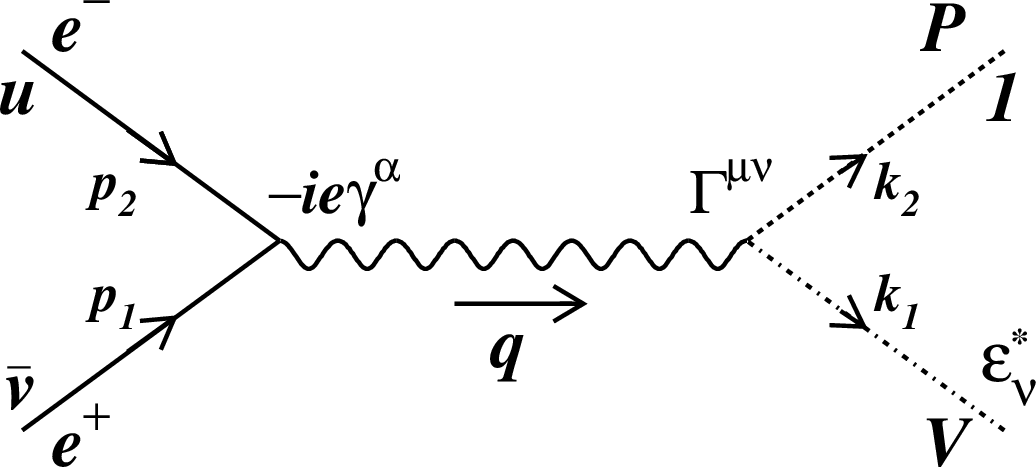}
\caption{\label{fig_pvfs}Feynman diagram of $\EE \to PV$ process.}
\end{figure}

The covariant vertex of $\EE \to PV$ process reads
\beq
\frac{1}{2} \epsilon_{\mu\nu\lambda\sigma}\partial^{\mu}V^{\nu}F^{\lambda\sigma} P ~.
\label{vtxpvfs}
\eeq
Expressing the vertex in momentum representation, we have
\beq
\begin{array}{rcl}
{\displaystyle \frac{1}{2}\epsilon_{\mu\nu\lambda\sigma}\partial^{\mu}V^{\nu}F^{\lambda\sigma}P }
    &=& {\displaystyle \frac{1}{2}\epsilon_{\mu\nu\lambda\sigma}\partial^{\mu} V^{\nu}(\partial^\lambda U^\sigma -\partial^\sigma U^\lambda) P } \\
    &=& {\displaystyle - \epsilon_{\mu\nu\lambda\sigma} k_1^{\mu} V^{\nu} k^\lambda U^\sigma P } \\
    &=& {\displaystyle - \epsilon_{\mu\nu\lambda\sigma} k^\lambda k_1^\sigma U^{\mu} V^{\nu} P } ~,
\end{array}
\label{vtxppxpn}
\eeq
from which the current is obtained
\beq
\Gamma^{\gamma P V}_{\mu\nu}=- e~ \epsilon_{\mu\nu\lambda\sigma}~ k^\lambda k_1^\sigma~.
\label{krtpvfs}
\eeq

According to Feynman rules and referring to Fig.~\ref{fig_pvfs}, the covariant amplitude reads
\beq
\begin{array}{rcl}
i\ampM&=& {\displaystyle \bar{v}(-ie\gamma^{\rho}) u \cdot \frac{-ig_{\rho \mu}}{q^2+i\epsilon}
       \cdot \Gamma^{\gamma P V}_{\mu \alpha}\epsilon^{*\alpha} P  }\\
    &=&{\displaystyle \frac{e^2}{q^2}(\bar{v}\gamma^{\mu} u)\cdot
       (\epsilon_{\mu\alpha\lambda\sigma}~ k^\lambda k_1^\sigma \epsilon^{*\alpha})~, }
\end{array}
\label{ampofpv}
\eeq
then
\beq
\begin{array}{rcl}
|\overline {\ampM }|^2 &=&{\displaystyle \frac{e^4}{q^4} [ p^{\mu}_1 p^{\nu}_2+p^{\nu}_1 p^{\mu}_2
 -g^{\mu\nu} ( p_1 \cdot p_2)] } \\
 & \times & {\displaystyle\sum\limits_{spins}
 [ \epsilon_{\mu\alpha\lambda\sigma}~ k^\lambda k_1^\sigma \epsilon^{*\alpha}  \cdot
   \epsilon_{\nu\beta\rho\tau}~ k^\rho k_1^\tau \epsilon^{\beta} ] }~.
\end{array}
\label{ttampofpva}
\eeq
First, it is noticed that
\beq
\sum\limits_{spins} \epsilon^{*\alpha} \epsilon^{\beta} =
-g^{\alpha \beta} +\frac{k_1^{\alpha} k_1^{\beta}}{m^2_V}~,
\label{sumforepsilon}
\eeq
where $m_V$ is the mass of vector meson $V$. Second, utilizing Eq.~\eref{twoepsilona}, it is obtained immediately
\beq
\begin{array}{l}
\epsilon_{\mu\alpha\lambda\sigma} \epsilon_{\nu\beta\rho\tau} \cdot
k^\lambda k_1^\sigma  k^\rho k_1^\tau k_1^{\alpha} k_1^{\beta} =\\
- \left|\begin{array}{cccc}
g_{\mu\nu}  & k_{1 \mu}   & k_{\rho}    & k_{1 \mu}   \\
k \cdot k_1 & k_1^2       & k \cdot k_1 & k_1^2       \\
k_{\nu}     & k \cdot k_1 & k^2         & k \cdot k_1 \\
k_{1 \nu}   & k_1^2       & k \cdot k_1 & k_1^2       \\
\end{array}\right|=0~,
\end{array}
\label{abtexln}
\eeq
since the second and fourth columns of the determinant are identical. Third, utilizing Eq.~\eref{twoepsilonb}, it is acquired finally
\beq
\begin{array}{rcl}
|\overline {\ampM }|^2 &=&{\displaystyle \frac{2 e^4}{q^4}
[ k \cdot k_1~( k \cdot p_1~ k_1 \cdot p_2+  k \cdot p_2~ k_1 \cdot p_1)  } \\
 &  & {\displaystyle k_1^2~ k \cdot p_1~ k \cdot p_2-  k^2~ k_1 \cdot p_1~ k_1 \cdot p_2 ] }~.
\end{array}
\label{ttampofpvb}
\eeq

Replacing the above terms with the corresponding values in CMS as given in Eqs.~\eref{pkdotxpna} and \eref{kpdotxpnb}, there results the expression
\beq
|\overline {\ampM }|^2= \frac{e^4}{2} |\vec{k}|^2 (1+\cos^2 \theta)~.
\label{ttampofpvc}
\eeq
Take into consideration the normalization factor $\xi=1/E^2$, we deduce the differential cross section
\beq
\frac{d\sigma_{PV} }{d\Omega} = \frac{\alpha^2 \beta^3}{8 s } \cdot (1+\cos^2 \theta)~.
\label{dffxnpp}
\eeq

\subsection{PA final state}
\begin{figure}[htb]
\includegraphics[width=7.cm]{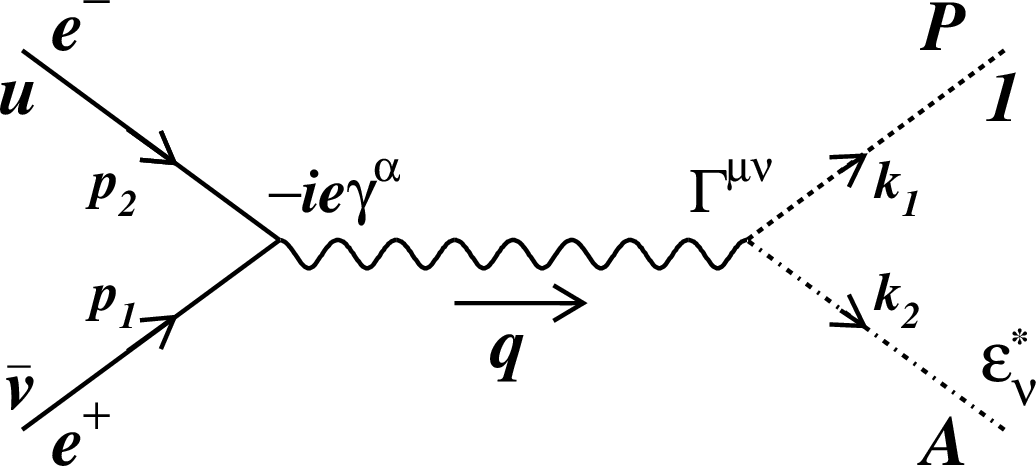}
\caption{\label{fig_pafs}Feynman diagram of $\EE \to PA$ process.}
\end{figure}

The covariant vertex of $\EE \to PA$ process reads
\beq
F_{\mu \nu } A^{\mu} \partial^{\nu} P ~.
\label{vtxpafs}
\eeq
Employing the momentum representation, we have
\beq
\begin{array}{rcl}
F_{\mu \nu } A^{\mu} \partial^{\nu} P
    &=& (\partial_{\mu} U_{\nu} - \partial_{\nu} U_{\mu} ) A^{\mu} \partial^{\nu} P \\
    &=& [g_{\mu \nu} (k \cdot k_1) - k_{1 \mu} k_{\nu} ] U^{\mu} A^{\nu} P~,
\end{array}
\label{vtxpaxpn}
\eeq
from which the current is obtained
\beq
\Gamma^{\gamma P A}_{\mu \nu}= e~ [g_{\mu \nu} (k \cdot k_1) - k_{1 \mu} k_{\nu} ]~.
\label{krtpafs}
\eeq
According to Feynman rules and referring to Fig.~\ref{fig_pafs}, the covariant amplitude reads
\beq
\begin{array}{rcl}
i\ampM&=& {\displaystyle \bar{v}(-ie\gamma^{\rho}) u \cdot \frac{-ig_{\rho \mu}}{q^2+i\epsilon}
       \cdot \Gamma^{\gamma P A}_{\mu \alpha} \epsilon^{*\alpha} P }\\
    &=&{\displaystyle \frac{e^2}{q^2}(\bar{v}\gamma^{\mu} u)
    [(k \cdot \epsilon^{*} ) k_{1\mu}-(k \cdot k_1) \epsilon^{*}_{\mu}]~},
\end{array}
\label{ampofpa}
\eeq
then
\beq
\begin{array}{rcl}
|\overline {\ampM }|^2 &=&{\displaystyle \frac{e^4}{q^4} [ p^{\mu}_1 p^{\nu}_2+p^{\nu}_1 p^{\mu}_2-g^{\mu\nu} (p_1 \cdot p_2)] } \\
  & \times & {\displaystyle\sum\limits_{spins}
 [(k \cdot \epsilon^{*} ) k_{1\mu}-(k \cdot k_1) \epsilon^{*}_{\mu}] }\\
  & \cdot &  [(k \cdot \epsilon ) k_{1\nu}-(k \cdot k_1) \epsilon_{\nu}] ~.
\end{array}
\label{ttampofpaa}
\eeq
Using the relation
\beq
\sum\limits_{spins} \epsilon^{*}_{\mu} \epsilon_{\nu} =
-g_{\mu \nu} +\frac{k_{1\mu} k_{1\nu}}{m^2_A}~,
\label{sumforepsilon}
\eeq
some calculations lead to
\beq
\begin{array}{l}
|\overline {\ampM }|^2={\displaystyle \frac{e^4}{q^4} \left\{ \left[-k^2+\frac{(k\cdot k_2)^2}{m_A^2} \right]
 [ 2(p_2\cdot k_1) (p_1 \cdot k_1) \right. } \\
   {\displaystyle -k_1^2(p_1 \cdot p_2) ]+(k\cdot k_1)^2 \left[(p_1\cdot p_2)
  +\frac{2(p_2 \cdot k_2)(p_1\cdot k_2)}{m_A^2} \right] } \\
   {\displaystyle -2(k\cdot k_1)\left[(p_1 \cdot k_1) \left( -(p_2 \cdot k)
  +(k\cdot k_2) \frac{(p_2\cdot k_2)}{m_A^2} \right) \right. } \\
   {\displaystyle +(p_2 \cdot k_1)\left(-(p_1\cdot k)+(k\cdot k_2)\frac{(p_1 \cdot k_2)}{m_A^2} \right) } \\
   {\displaystyle \left.\left. -(p_1 \cdot p_2)\left(-(k\cdot k_1)
   +(k\cdot k_2)\frac{(k_1\cdot k_2)}{m_A^2}\right)\right] \right \} }
\end{array}
\label{ttampofpab}
\eeq

Applying the corresponding values in Eqs.~\eref{pkdotxpna} and \eref{kpdotxpnb}, it follows that
\beq
\begin{array}{rl}
|\overline {\ampM }|^2&{\displaystyle  = e^4 m_P^2 \left\{
1+\left(1+\frac{s}{m_A^2}\right)\frac{|\vec{k}|^2}{2m_P^2} \right. } \\
 &{\displaystyle \left. + \left(1-\frac{s}{m_A^2}\right)\frac{|\vec{k}|^2}{2m_P^2} \cos^2 \theta \right\} }~.
\end{array}
\label{ttampofpab}
\eeq
With taking the normalization factor $\xi=1/E^2$ into account, the differential cross section has the form
\beq
\begin{array}{rl}
{\displaystyle \frac{d\sigma_{PA}}{d\Omega}}&{\displaystyle  = \frac{\alpha^2\beta}{8 s}\frac{m_P^2}{E^2} \left\{
1+\left(1+\frac{s}{m_A^2}\right)\frac{|\vec{k}|^2}{2m_P^2} \right. } \\
 &{\displaystyle \left. + \left(1-\frac{s}{m_A^2}\right)\frac{|\vec{k}|^2}{2m_P^2} \cos^2 \theta \right\} }~.
\end{array}
\label{dffxnpa}
\eeq
Upon integrating the angular variables, one gets the total cross section
\beq
\sigma_{PA} (s) =\frac{\pi\alpha^2\beta}{2 s} \frac{m_P^2}{E^2}
\left[1 + \frac{2}{3}\left( 2+ \frac{s}{m_A^2}\right)\frac{|\vec{k}|^2}{2m_P^2} \right]~.
\label{ttxnpa}
\eeq

\subsection{PT final state}
\begin{figure}[htb]
\includegraphics[width=7.cm]{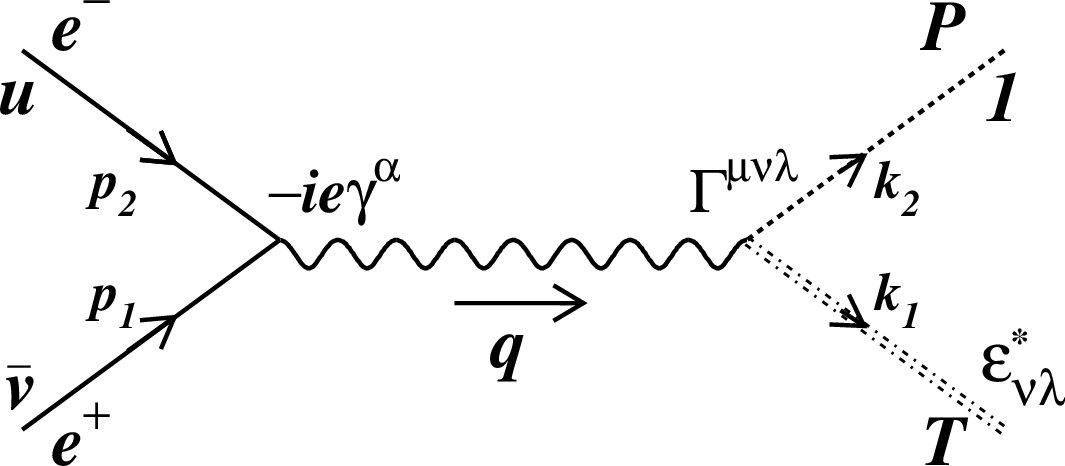}
\caption{\label{fig_ptfs}Feynman diagram of $\EE \to PT$ process.}
\end{figure}

The covariant vertex of $\EE  \to PT$ process reads
\beq
\frac{1}{2}\epsilon_{\mu\nu\lambda\sigma}T^{\mu\delta}F^{\nu\lambda}\partial_\delta \partial^\sigma P~.
\label{vtxptfs}
\eeq
Employing the momentum representation, we have
\beq
\begin{array}{rl}
{\displaystyle
\frac{1}{2}\epsilon_{\mu\nu\lambda\sigma}T^{\mu\delta}F^{\nu\lambda}\partial_\delta \partial^\sigma P }
    &={\displaystyle \frac{1}{2}\epsilon_{\mu\nu\lambda\sigma}T^{\mu\delta}(\partial^\nu U^\lambda-\partial^\lambda U^\nu)\partial_\delta \partial^\sigma P } \\
    &={\displaystyle -i \epsilon_{\mu\nu\lambda\sigma}T^{\mu\delta}k^\nu U^\lambda  k_{2\delta}k_2^\sigma P } \\
    &={\displaystyle i\epsilon_{\mu\nu\lambda\sigma}k^\nu k_{2\delta}k_2^\sigma  U^{\mu}T^{\lambda\delta}P } ~,
\end{array}
\label{vtxptxpn}
\eeq
from which the current is obtained
\beq
\Gamma^{\gamma P T}_{\mu\lambda\delta}=i~e~\epsilon_{\mu\nu\lambda\sigma}k^\nu  k_{2\delta}k_2^\sigma~.
\label{krtptfs}
\eeq

According to Feynman rules and referring to Fig.~\ref{fig_ptfs}, the covariant amplitude reads
\beq
\begin{array}{rcl}
i\ampM&=& {\displaystyle \bar{v}(-ie\gamma^{\rho}) u \cdot \frac{-ig_{\rho \mu}}{q^2+i\epsilon}
       \cdot \Gamma^{\gamma P T}_{\mu\lambda\delta} \epsilon^{*\lambda\delta} P  }\\
    &=&{\displaystyle -\frac{i e^2}{q^2}(\bar{v}\gamma^{\mu} u)\cdot
       (\epsilon_{\mu\alpha\lambda\sigma}~ k^\alpha k_{2\delta}k_2^\sigma \epsilon^{*\lambda\delta})~, }
\end{array}
\label{ampofpv}
\eeq
then
\beq
\begin{array}{rl}
|\overline {\ampM }|^2 &={\displaystyle \frac{e^4}{q^4} [ p^{\mu}_1 p^{\nu}_2+p^{\nu}_1 p^{\mu}_2
 -g^{\mu\nu} ( p_1 \cdot p_2)] } \\
 \times & {\displaystyle\sum\limits_{spins}
 [ \epsilon_{\mu\alpha\lambda\sigma}~ k^\alpha k_{2\delta}k_2^\sigma \epsilon^{*\lambda\delta}  \cdot
   \epsilon_{\nu\beta\rho\tau}~ k^\beta k_{2\eta}k_2^\tau \epsilon^{\rho\eta} ] }~.
\end{array}
\label{ttampofpta}
\eeq
It is noticed that~\cite{huangsz2003,huangsz2005}
\beq
\begin{array}{rl}
\sum\limits_{spins} \epsilon^{*\lambda\delta}\epsilon^{\lambda^\prime \delta^\prime}
&={\displaystyle  \frac{1}{2}\left(g^{\lambda\lambda^\prime}-\frac{k^{\lambda}k^{\lambda^\prime}}{m_T^2} \right)
\left(g^{\delta\delta^\prime}-\frac{k^{\delta}k^{\delta^\prime}}{m_T^2} \right) } \\
&+{\displaystyle \frac{1}{2} \left(g^{\lambda\delta^\prime}-\frac{k^{\lambda}k^{\delta^\prime}}{m_T^2} \right)
\left( g^{\delta\lambda^\prime}-\frac{k^{\delta}k^{\lambda^\prime}}{m_T^2} \right) }  \\
&{\displaystyle  -\frac{1}{3}\left(g^{\lambda\delta}-\frac{k^{\lambda}k^{\delta}}{m_T^2} \right)
\left(g^{\lambda^\prime \delta^\prime}-\frac{k^{\lambda^\prime}k^{\delta^\prime}}{m_T^2} \right)~, }
\end{array}
\label{smofeslnvt}
\eeq
where $m_T$ is the mass of tensor meson $T$. Utilizing the technique in analogy to that for $PV$ final state,  it is finally acquired
\beq
\begin{array}{l}
|\overline {\ampM }|^2 ={\displaystyle \frac{e^4}{q^4 m_T^2} \times } \\
\{ k_1^2 (p_1\cdot p_2)^2- (p_1\cdot p_2)[ (p_1 \cdot k_1)^2 + (p_2\cdot k_1)^2 \} \\
\cdot \{ -(p_1 \cdot k_1)^2 -(p_2 \cdot k_1)^2 -(p_1 \cdot k_1) [p_2 \cdot k_1 +m_T^2] \\
+ k_1^2 [2 p_1 \cdot k_1 +2 p_2 \cdot k_1 +m_T^2]  \\
-2 m_T^2 p_2 \cdot k_1 + 2 m_T^2 p_1 \cdot p_2 -k_1^4  \}~.
\end{array}
\label{ttampofptb}
\eeq

Recalling the corresponding values in Eqs.~\eref{pkdotxpna} and \eref{kpdotxpnb}, it yields
\beq
|\overline {\ampM }|^2= \frac{e^4 E^2 |\vec{k}|^4}{m_T^2}  (1+\cos^2 \theta)~.
\label{ttampofpvc}
\eeq
Taking into account the normalization factor $\xi=1/E^4$,  we conclude that
\beq
\frac{d\sigma_{PT} }{d\Omega} = \frac{\alpha^2 \beta^5}{4 s }\cdot
\frac{E^2}{m_T^2 } \cdot (1+\cos^2 \theta)~.
\label{dffxnpp}
\eeq

\begin{table*}[bth]
\caption{\label{decayampmmtx}Cross sections and relevant information of $\EE \to M_1 M_2$ process, where $M_1$ can be $S$ and $P$, while $M_2$ can be $S$, $P$, $V$, $A$, and $T$. For the current, the common factor $e$ has been suppressed. }
\begin{ruledtabular}
\begin{tabular}{lllll}
Final  & Vertex         & Current     & Angular distribution & Total section  \\
state  &                &             &                      &                \\   \hline
$S S^{\prime}$  & $F_{\mu \nu } \partial^{\mu} S \partial^{\nu} S^{\prime} $
                        & $ i [(k \cdot k_1 ) k_{2\mu}-(k \cdot k_2) k_{1\mu} ] $
                        & $1-\cos^2 \theta$
                        & ${\displaystyle \frac{4 \pi \alpha^2 \beta^3}{3 s } }$   \\
$S V$           & $F_{\mu \nu } V^{\mu \nu } S$
                        & $2[k_{1 \mu} k_{\nu} - g_{\mu \nu} (k \cdot k_1)]$
                        & ${\displaystyle 1+ \frac{|\vec{k}|^2}{2m_V^2}+\frac{|\vec{k}|^2}{2 m_V^2} \cos^2 \theta }$
                        & ${\displaystyle \frac{4 \pi \alpha^2 \beta}{3 s } \left( 3 + \frac{4|\vec{k}|^2}{2m_V^2} \right) }$ \\
$S A$           & ${\displaystyle \frac{1}{2} \epsilon_{\mu\nu\lambda\sigma} F_{\mu \nu } \partial^{\lambda} S A^{\sigma} }$           & $-\epsilon_{\mu\nu\lambda\sigma}~ k^\lambda k_1^\sigma$
                        & $1+\cos^2 \theta$
                        & ${\displaystyle \frac{2 \pi \alpha^2 \beta^3}{3 s } }$ \\
$S T$           & $F_{\mu \nu } T^{\mu\lambda} \partial_\lambda \partial^\nu S $
                        & $i[k_{\mu} k_{1 \nu} k_{1 \lambda} - g_{\mu \nu} k_{1 \lambda} (k \cdot k_1) ]$
 & $\begin{array}{l} {\displaystyle \left[ 7+\left(3+\frac{4s}{m_T^2}\right)\frac{|\vec{k}|^2}{m_S^2} \right] - }\\
 {\displaystyle \left[1+ \left(-3+\frac{4s}{m_T^2}\right)\frac{|\vec{k}|^2}{m_S^2} \right]\cos^2 \theta } \end{array} $
                        & ${\displaystyle \frac{4\pi\alpha^2\beta}{9 s} \frac{m_S^2}{m_T^2}
\left[5 + \left( 3+\frac{2s}{m_T^2}\right)\frac{|\vec{k}|^2}{m_S^2} \right]  }$ \\
$P P^{\prime}$  & $F_{\mu \nu } \partial^{\mu} P \partial^{\nu} \Pp $
                        & $ i [(k \cdot k_1 ) k_{2\mu}-(k \cdot k_2) k_{1\mu} ] $
                        & $1-\cos^2 \theta$
                        & ${\displaystyle \frac{4 \pi \alpha^2 \beta^3}{3 s } }$   \\
$P V$           & ${\displaystyle \frac{1}{2} \epsilon_{\mu\nu\lambda\sigma} \partial^{\mu} V^{\nu} F^{\lambda\sigma} P }$
                        & $- \epsilon_{\mu\nu\lambda\sigma}~ k^\lambda k_1^\sigma$
                        & $1+\cos^2 \theta$
                        & ${\displaystyle  \frac{2 \pi \alpha^2 \beta^3}{3 s } }$  \\
$P A$           & $F_{\mu \nu }A^{\mu} \partial^{\nu} P$
                        & $ [g_{\mu \nu} (k \cdot k_1) - k_{1 \mu} k_{\nu} ]$
              & $\begin{array}{l} {\displaystyle 1+\left(1+\frac{s}{m_A^2}\right)\frac{|\vec{k}|^2}{2m_P^2} + }\\
            {\displaystyle \left(1-\frac{s}{m_A^2}\right)\frac{|\vec{k}|^2}{2m_P^2} \cos^2 \theta  } \end{array} $
                        & ${\displaystyle \frac{\pi\alpha^2\beta}{2s} \frac{m_P^2}{E^2}
\left[1 + \frac{2}{3}\left( 2+ \frac{s}{m_A^2}\right)\frac{|\vec{k}|^2}{2m_P^2} \right]  }$  \\
$P T$           & ${\displaystyle  \frac{1}{2} \epsilon_{\mu\nu\lambda\sigma} T^{\mu\delta} F^{\nu\lambda}\partial_\delta \partial^\sigma P }$
                        & $i\epsilon_{\mu\nu\lambda\sigma}k^\nu k_{2\delta}k_2^\sigma$
                        & $1+\cos^2 \theta$
                        & ${\displaystyle \frac{4 \pi \alpha^2 \beta^5}{3 s }\cdot \frac{E^2}{m_T^2 }  }$
\end{tabular}
\end{ruledtabular}
\end{table*}

\subsection{Comment }
In the proceeding section, each mode has been accorded separte treatment. Observed that for the same kind of particle, the corresponding strategy is similar. Therefore, the aforementioned technique for above final states can be extended easily to other final states. Also performed are the analogous calculations for the other four kinds of final states, that is scalar-scalar (SS), scalar-vector (SV), scalar-axial-vector (SA), and scalar-tensor (ST). All results are tabulated in Table~\ref{decayampmmtx}.

\section{Parametrization of charmonium decay}\label{xct_chamdk}

In this section, we are to discuss the parametrization of charmonium decay, and focus on the $V P$ final state. The reason lies in that not only because the $V P$ mode is one of two typical decay modes~\cite{moxh2024}, but also because all $V P$ final states have experimental measurements that can be utilized for the data analysis in the future.

\subsection{Born section}
According to the result of the previous section, the total cross section for $\EE \to V P$ process reads
\beq
\sigma_{VP} (s) = \frac{2 \pi \alpha^2 \beta^3}{3 s} \ffkt (s) ~,
\label{ttxknvp}
\eeq
where $\ffkt (s)=f^2_{VP}(q^2)~(q^2=s)$, and $f_{VP}(q^2)$ is the form factor that has been discussed in
subsection~\ref{xct_xnppfs}. Usually, $\ffkt (s)$ is called a form factor as well, in which the internal feature due to quarks has been subsumed.

For $\EE$ colliding experiments, besides the continuum electromagnetic process~\cite{wangp03hepnp}
$$ 
\EE \rightarrow \gamma^* \to VP~,
$$ 
there is a most important resonance process
$$ 
\EE \rightarrow \psi \to VP~.
$$ 

\begin{figure}[b]
\begin{minipage}{8cm}
\includegraphics[width=3.25cm,height=2.5cm]{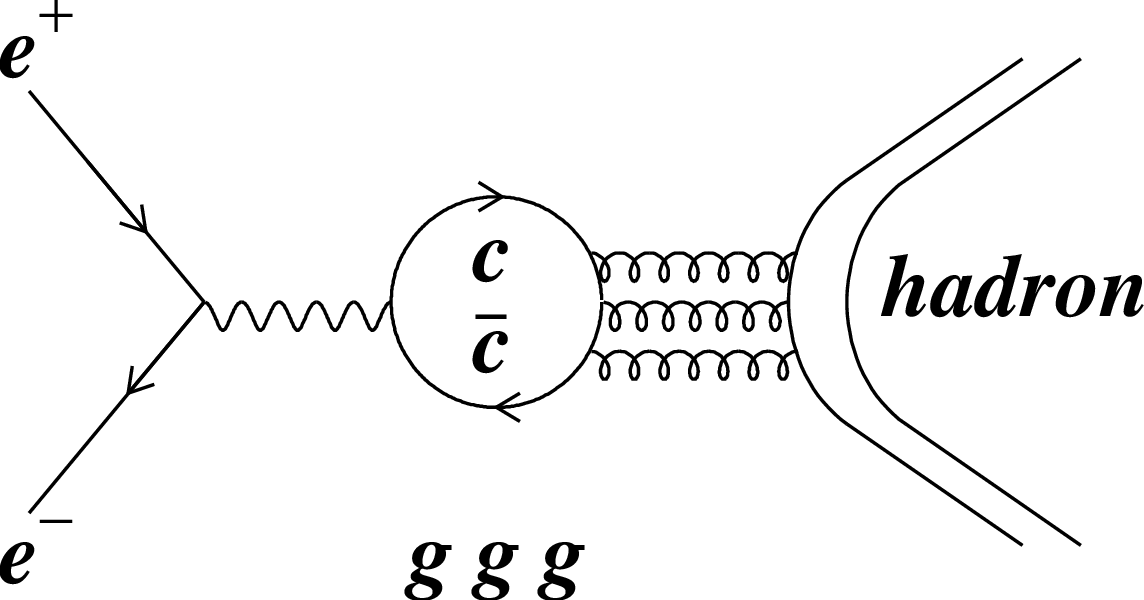}
\hskip 0.5cm
\includegraphics[width=3.25cm,height=2.5cm]{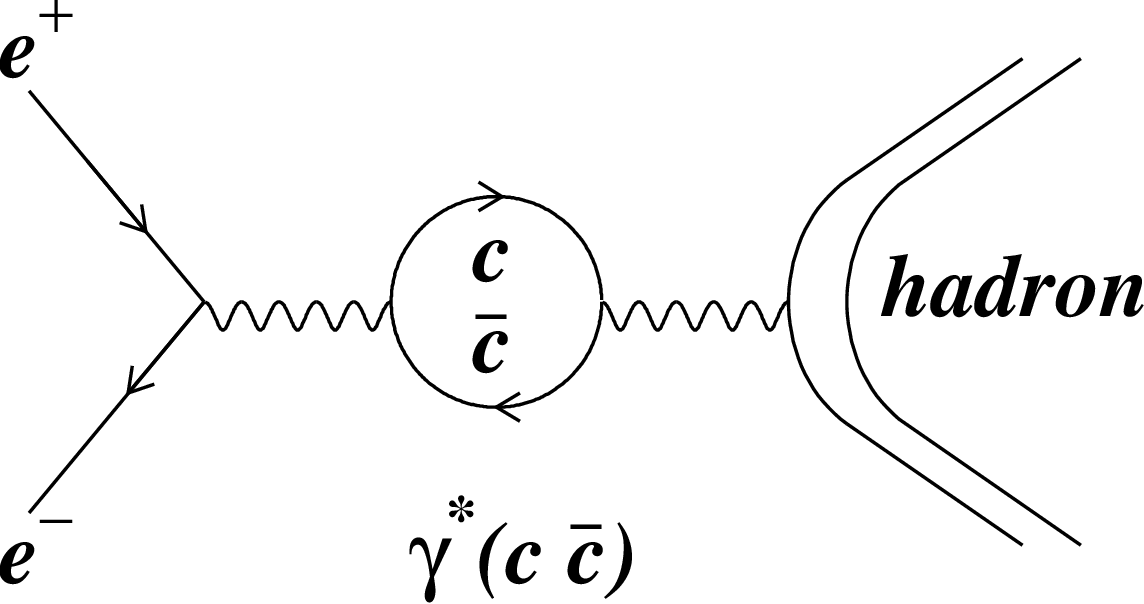}
\end{minipage}
\begin{minipage}{8cm}
\center
\includegraphics[width=3.5cm,height=2.5cm]{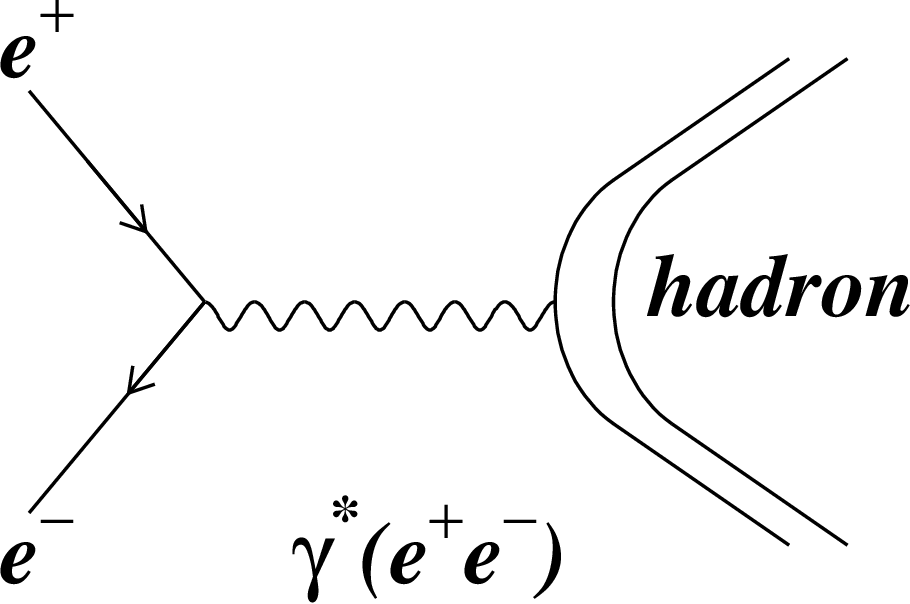}
\end{minipage}
\caption{\label{threefymn} The three classes of diagrams of
$\EE\rightarrow light\, \, hadrons$ at charmonium resonance. The
charmonium state is represented by a charm quark loop.}
\end{figure}

For the resonance process, $\psi$ decays to $VP$ can through three-gluon or one-photon. Therefore, three are totally three amplitudes for a certain final state, as shown in Fig.~\ref{threefymn}, where the $hadron$ indicates the $VP$ final state for our study here. The Born order cross section for a final state $f$ is~\cite{rudaz,wymcgam,Wang:2005sk,wymphase,wymogpiapp}
\beq
\sigma_{Born} = \frac{2 \pi \alpha^2 \beta^3}{3 s} |A_f(s)|^2 \cdot \ffkt (s) ~.
\label{Born}
\eeq
Here $f$ denotes one of VP final states, and
\beq
A_f(s) = \ag(s)+\aga(s)+\ac(s) ~,
\label{amptot}
\eeq
which consists of three kinds of amplitudes corresponding to (a) the strong interaction [$\ag(s)$] presumably through three-gluon annihilation, (b) the electromagnetic interaction [$\aga(s)$]
through the annihilation of $c\overline{c}$ pair into a virtual photon, and (c) the electromagnetic interaction [$\ac(s)$] due to one-photon continuum process. Specially, the amplitudes have the forms~:

\beq
\ac(s)=Y_f ~,
\label{ampac}
\eeq
\beq
\aga(s)=Y_f \cdot B(s) ~,
\label{ampap}
\eeq
\beq
\ag(s)=X_f \cdot B(s) ~,
\label{ampag}
\eeq
with the definition
\beq
B(s) \equiv \frac{3\sqrt{s}\Gamma_{ee}/\alpha}{s-M^2+iM\Gamma_t}~~,
\label{defbsfcn}
\eeq
where $\alpha$ is the QED fine structure constant; $M$ and $\Gamma_t$ are the mass and the total width of the $\psp$ or $\jpsi$; $\Gamma_{ee}$ is the partial width to $\EE$. So far as $X_f$ and $Y_f$ are concerned, the symmetry analysis results will be used to figure out their concrete forms.

\subsection{Symmetry analysis}
By virtue of symmetry analysis, a systematic parametrization scheme has been established through a series of papers~\cite{moxh2022,moxh2023,moxh2024,moxh2025}. Here we merely recapitulate the main ideas and present the results relevant to the $VP$ mode.

It is well known that in the $\EE$ collider experiment, the initial state is obviously flavorless, then the final state must be flavor singlet. If only the Okubo-Zweig-Iizuka (OZI) rule suppressed processes are considered, the final states merely involve light quarks, that is $u, d, s$ quarks. Therefore, $SU(3)$ group is employed for symmetry analysis. The key rule herein is the so-call ``flavor singlet principle'' that determines what kinds of terms are permitted in the effective interaction Hamiltonian. Resorting to the perturbation language, the Hamiltonian is written as
\beq
\Heff = H_0 + \Delta H~,
\label{perturbaionhmtn}
\eeq
where $H_0$ is the symmetry conserved term and $\Delta H$ the symmetry breaking term, which is generally small compare to $H_0$. In light of group representation theory, the product of two multiplets, say ${\mathbf n}$ and ${\mathbf m}$, can be decomposed into a series of irreducible representations, that is
\beq
{\mathbf n} \otimes {\mathbf m} = {\mathbf l_1} \oplus {\mathbf l_2} \oplus \cdots \oplus {\mathbf
l_k}~.
\label{dcpsmoftwomtplt}
\eeq
The singlet principle requires that among the ${\mathbf l_j} (j=1, \cdots, k)$, only the singlet term, i.e., ${\mathbf l_j}={\mathbf 1}$ for certain $j$, is allowed in the Hamiltonian. Since this term is obviously $SU(3)$ invariant, it is called the symmetry conserved term, and denoted by $H_0$.

As far as the $SU(3)$-breaking effect is concerned, they are treated as a ``spurion'' octet, then the favor singlet principle is used to pin down the breaking term in Hamiltonian. There are totally three kinds of effects~\cite{moxh2025}, that is the strong breaking effect, the electromagnetic breaking effect, and the breaking effect due to the magnetic momentum of quarks, which can be expressed in tensor forms as
\beq
({\mathbf S}_{m})^{i}_{j} = g_m \delta^i_3 \delta^3_j~,
\label{smdefn}
\eeq
\beq
({\mathbf S}_{e})^{i}_{j} = g_e \delta^i_1 \delta^1_j~,
\label{sedefn}
\eeq
and
\beq
({\mathbf S}_{\mu})^{i}_{j} = g_{\mu} \delta^i_2 \delta^2_j~.
\label{sudefn}
\eeq
It is worthy of noticing that these three breaking effects de facto fully exhaust the possible symmetry of elementary representation. From a pure viewpoint of group theory, three spin-symmetries are not independent. If the coefficients of all three breaking effect are exactly the same, the synthetic effect will be zero. Nevertheless,
the physical original of these breaking effects are fairly distinct, therefore the coefficients of them should be different.

It is a well known fact that for a baryon, the particle and its antiparticle are in different multiplets; but for a meson, the particle and its antiparticle are in the same multiplet. Under such a condition, the introduction of a generalized inherent ${C}$-parity is crucial for the determination of the effective Hamiltonian. Herein, it is assumed that the eigenvalue of the generalized inherent ${C}$-parity is equal to that of the neutral particle in the multiplet. For two octet meson final states, denoted respectively by $O_1$ and $O_2$, defined are the following terms, which may be allowed or forbidden in the effective Hamiltonian:
\beq
[O_1 O_2]_0 = (O_1)^i_j (O_2)^j_i~~,
\eeq
\beq
([O_1 O_2]_f )^i_j = (O_1)^i_k (O_2)^k_j -(O_1)^k_j (O_2)^i_k~~,
\eeq
and
\beq
([O_1 O_2]_d )^i_j = (O_1)^i_k (O_2)^k_j +(O_1)^k_j (O_2)^i_k
-\frac{2}{3} \delta^i_j \cdot (O_1)^i_j (O_2)^j_i~~.
\eeq
Under parity transformation, $\hat{C} [O_1 O_2]_{x} \to \xi_{x} [O_1 O_2]_{x}$, where $x=0,d,f$, that is $\xi_{0}=+1,\xi_{d}=+1,\xi_{f}=-1$. In addition, $\hat{C} O_i \to \eta_{O_i} O_i, (i=1,2)$, synthetically,
\beq
\hat{C}~[O_1 O_2]_{x}  = \eta_{O_1} \eta_{O_2} \xi_x [O_1 O_2]_{x}~,  \\
\label{ctfmnforokt}
\eeq
At the same time for the initial state of $\psi$, $\hat{C}~\psi  = \eta_{\psi} \psi$. Then the term $[O_1 O_2]_{x}$ is allowed in the effective Hamiltonian as long as $\eta_{\psi}=-1=\eta_{O_1} \eta_{O_2} \xi_x$, otherwise, it is forbidden. With this criterion, it is easy to figure out what kind of terms can be adopted in the effective Hamiltonian for various kinds of final states~\cite{moxh2024}. As a matter of fact, there exist merely two types of Hamiltonian forms. One contains both $[O_1 O_2]_{0}$ and $[O_1 O_2]_{d}$ terms, while the other contains only $[O_1 O_2]_{f}$ term, that is
\beq
\left.\begin{array}{rl}
\Heff^{O_1 O_2}= & \gz \cdot [O_1 O_2]_0 + g_m \cdot ([O_1 O_2]_d )^3_3 \\
        & + g_e \cdot ([O_1 O_2]_d )^1_1 + g_{\mu} \cdot ([O_1 O_2]_d )^2_2 ~,
\end{array}\right.~~
\label{effhmtvptype}
\eeq
or
\beq
\Heff^{O_1 O_2} =  g_m \cdot ([O_1 O_2]_f )^3_3  + g_e \cdot ([O_1 O_2]_f )^1_1
 + g_{\mu} \cdot ([O_1 O_2]_f )^2_2 ~.
\label{effhmtpptype}
\eeq
If $O_1=V$ (Vector) and $O_2=P$ (Pseduoscalar) in Eq.~\eref{effhmtvptype}, the corresponding parametrization for $VP$ final state can be obtained.

\begin{table*}[bth]
\caption{\label{vpmsnfmmix}Amplitude parametrization form for decays of the $\psp$ or $\jpsi$ into $V~P$ final states. The mixing between octet and singlet are taken into account according to Eqs.~\eref{mixpsmsnrv} and \eref{mixvtmsnrv}. The shorthand symbols are defined as $s_{\alpha}=\sin \theta_{\alpha}$ and $c_{\alpha}=\cos \theta_{\alpha}$ $(\alpha=V,P)$. }
\begin{ruledtabular}
\begin{tabular}{ccccc}
Decay mode    &\multicolumn{4}{c}{Coupling constant} \\ \cline{2-5}
$\psi \to X$  &$g_0$    &   $g_m$  & $g_e$ &  $g_{\mu}$ \\ \hline
$\roz \piz$ &  1  &$-2/3$ &$ 1/3$ &$ 1/3$ \\
$\rop \pim$ &  1  &$-2/3$ &$ 1/3$ &$ 1/3$ \\
$\rom \pip$ &  1  &$-2/3$ &$ 1/3$ &$ 1/3$ \\
$\kstp\kam$ &  1  &$ 1/3$ &$ 1/3$ &$-2/3$ \\
$\kstm\kap$ &  1  &$ 1/3$ &$ 1/3$ &$-2/3$ \\
$\kstz\kazb$&  1  &$ 1/3$ &$-2/3$ &$ 1/3$ \\
$\kstzb\kaz$&  1  &$ 1/3$ &$-2/3$ &$ 1/3$ \\
$\phi\eta$  &$c_V c_P+s_V s_P$
  &$\frac{2}{3} c_V c_P+\frac{2\sqrt{2}}{3}(c_V s_P +s_V c_P)$
  &$-\frac{1}{3} c_V c_P-\frac{\sqrt{2}}{3}(c_V s_P +s_V c_P)$
  &$-\frac{1}{3} c_V c_P-\frac{\sqrt{2}}{3}(c_V s_P +s_V c_P)$ \\
$\phi\etap$ &$c_V s_P-s_V c_P$
  &$\frac{2}{3} c_V s_P-\frac{2\sqrt{2}}{3}(c_V c_P -s_V s_P)$
  &$-\frac{1}{3} c_V s_P+\frac{\sqrt{2}}{3}(c_V c_P -s_V s_P)$
  &$-\frac{1}{3} c_V s_P+\frac{\sqrt{2}}{3}(c_V c_P -s_V s_P)$ \\
$\omega\eta$&$s_V c_P-c_V s_P$
  &$\frac{2}{3} s_V c_P-\frac{2\sqrt{2}}{3}(c_V c_P -s_V s_P)$
  &$-\frac{1}{3} s_V c_P+\frac{\sqrt{2}}{3}(c_V c_P -s_V s_P)$
  &$-\frac{1}{3} s_V c_P+\frac{\sqrt{2}}{3}(c_V c_P -s_V s_P)$ \\
$\omega\etap$&$s_V s_P+c_V c_P$
  &$\frac{2}{3} s_V s_P-\frac{2\sqrt{2}}{3}(c_V s_P +s_V c_P)$
  &$-\frac{1}{3} s_V s_P+\frac{\sqrt{2}}{3}(c_V s_P +s_V c_P)$
  &$-\frac{1}{3} s_V s_P+\frac{\sqrt{2}}{3}(c_V s_P +s_V c_P)$ \\
$\roz\eta$  &   0 &0 &$\sqrt{\frac{1}{3}}c_P-\sqrt{\frac{2}{3}}s_P$ &$-(\sqrt{\frac{1}{3}}c_P-\sqrt{\frac{2}{3}}s_P)$\\
$\roz\etap$ &   0 &0 &$\sqrt{\frac{1}{3}}s_P+\sqrt{\frac{2}{3}}c_P$ &$-(\sqrt{\frac{1}{3}}s_P+\sqrt{\frac{2}{3}}c_P)$\\
$\phi\piz$  &   0 &0 &$\sqrt{\frac{1}{3}}c_V-\sqrt{\frac{2}{3}}s_V$ &$-(\sqrt{\frac{1}{3}}c_V-\sqrt{\frac{2}{3}}s_V)$\\
$\omega\piz$&   0 &0 &$\sqrt{\frac{1}{3}}s_V+\sqrt{\frac{2}{3}}c_V$ &$-(\sqrt{\frac{1}{3}}s_V+\sqrt{\frac{2}{3}}c_V)$\\
\end{tabular}
\end{ruledtabular}
\end{table*}

In above analysis, all particles are assumed to be pure octet components, including the mesons $\omega$ and $\eta$, but the real or observed $\omega$ and $\eta$ are actually the mixing of pure octet and singlet components with a certain mixing angle. According to PDG~\cite{pdg2020},
\beq
\left.\begin{array}{rcl}
\phi   &=&\omega^8~\cos \theta_V -\omega^1~\sin \theta_V~,   \\
\omega &=&\omega^8~\sin \theta_V +\omega^1~\cos \theta_V~,
\end{array}\right.
\label{mixvtmsn}
\eeq
or its reverse
\beq
\left.\begin{array}{rcl}
\omega^8  &=&\phi~\cos \theta_V  + \omega~\sin \theta_V~,   \\
\omega^1 &=&-\phi~\sin \theta_V + \omega~\cos \theta_V~;
\end{array}\right.
\label{mixvtmsnrv}
\eeq
and
\beq
\left.\begin{array}{rcl}
\eta   &=&\eta^8~\cos \theta_P -\eta^1~\sin \theta_P~,   \\
\etap  &=&\eta^8~\sin \theta_P +\eta^1~\cos \theta_P~,
\end{array}\right.
\label{mixpsmsn}
\eeq
or its reverse
\beq
\left.\begin{array}{rcl}
\eta^8  &=&\eta~\cos \theta_P  + \etap~\sin \theta_P~,   \\
\eta^1 &=&-\eta~\sin \theta_P  + \etap~\cos \theta_P~.
\end{array}\right.
\label{mixpsmsnrv}
\eeq
In above equations, superscribe $8$ indicates the unitary octet component and superscribe $1$ the unitary singlet component. With the mixing angles being taken into account, the final parametrization can be acquired and listed in Table~\ref{vpmsnfmmix}.

Two remarks are in order here. First, the possible admixture of quarkonium with gluonium states is not taken into account for our study in this paper~\cite{moxh2025}. Second, the nonet symmetry is assumed. Therefore, instead of meson octets, actually adopted in Eq.~\eref{effhmtvptype} are meson nonets, which can be expressed as
\beq
{\mathbf V_N}=
\left(\begin{array}{ccc}
V^1_1   & \rop    & \kstp    \\
\rom    & V^2_2   & \kstz    \\
\kstm   & \kstzb  &  V^3_3
\end{array}\right)~~,
\label{notvtmsn}
\eeq
with
\beq
\left.\begin{array}{rcl}
V^1_1&=&\roz/\sqrt{2}+\omega^8/\sqrt{6} +\omega^1/\sqrt{3}~,    \\
V^2_2&=& -\roz/\sqrt{2}+\omega^8/\sqrt{6} +\omega^1/\sqrt{3}~,    \\
V^3_3&=& -2 \omega^8/\sqrt{6}+\omega^1/\sqrt{3}~;
\end{array}\right.
\eeq
and
\beq
{\mathbf P_N}=
\left(\begin{array}{ccc}
P^1_1    & \pip         & \kap    \\
\pim     & P^2_2  & \kaz    \\
\kam     & \kazb  & P^3_3
\end{array}\right)~~,
\label{notpsmsn}
\eeq
with
\beq
\left.\begin{array}{rcl}
P^1_1&=&\piz/\sqrt{2}+\eta^8/\sqrt{6} +\eta^1/\sqrt{3} ~,   \\
P^2_2&=& -\piz/\sqrt{2}+\eta^8/\sqrt{6} +\eta^1/\sqrt{3}~, \\
P^3_3&=& -2 \eta^8/\sqrt{6} +\eta^1/\sqrt{3}~.
\end{array}\right.
\eeq

Now we discuss the forms of $X_f$ and $Y_f$. As displayed in Fig.~\ref{threefymn}, $Y_f$ is related to the electromagnetic interaction. Based on the analysis of Ref.~\cite{moxh2022}, $g_e$ is the U-spin conversed coefficient, which is relevant to the electromagnetic interaction, therefore, it can be ascribed to $Y_f$. All other coefficients are strong interaction related, and can be ascribed to $X_f$. As the functions of interaction coefficients and mixing angles, the special forms of $X_f$ and $Y_f$ can be obtained from Table~\ref{vpmsnfmmix}, viz.
\beq
Y=Y(g_e, s_P, c_P, s_V, c_V)~,
\label{defy}
\eeq
\beq
X=X(\gz, g_m, g_{\mu},s_P, c_P, s_V, c_V) e^{i\varphi}~.
\label{defx}
\eeq
By virtue of Eq.~\eref{mixvtmsn} and Eq.~\eref{mixpsmsn}, $s_P=\sin \theta_P,~c_P=\cos \theta_P$ and $s_V=\sin \theta_V,~c_V=\cos \theta_V$, where $\theta_P$ and $\theta_V$ are respectively the mixing angle of pseudoscalar meson between $\eta$ and $\etap$, and that of vector meson between $\phi$ and $\omega$. The $\varphi$ is relative phase angle between the strong interaction and the electromagnetic interactions.
As examples, for the $\rho \pi$ final state,
\beq
\left.\begin{array}{rcl}
X_{\rho \pi}&=&{\displaystyle \gz-\frac{2}{3} \cdot g_m +\frac{1}{3} \cdot g_{\mu}  }~,   \\
Y_{\rho \pi}&=&{\displaystyle  \frac{1}{3} \cdot g_e }~;
\end{array}\right.
\eeq
for the $\omega\piz$ final state,
\beq
\left.\begin{array}{rcl}
X_{\omega\piz}&=&{\displaystyle  -(\sqrt{\frac{1}{3}}s_V+\sqrt{\frac{2}{3}}c_V) \cdot g_{\mu} } ~,   \\
Y_{\omega\piz}&=&{\displaystyle  (\sqrt{\frac{1}{3}}s_V+\sqrt{\frac{2}{3}}c_V) \cdot g_e }~.
\end{array}\right.
\eeq

In principle, the parameters $\gz$, $g_m$, $g_e$, and $g_{\mu}$ could be complex arguments, each with a magnitude together with a phase. Conventionally, it is assumed that there is not relative phases among the strong-originated coefficients amplitudes $\gz$, $g_m$, $g_{\mu}$, and the sole phase [denoted by $\varphi$ in Eq.~\eref{defx} ] is between the strong and electromagnetic interactions, that is between $X$ and $Y$, as indicated in Eqs.~\eref{defy} and \eref{defx}, where $\gz$, $g_m$, $g_e$, and $g_{\mu}$ are actually treated as real numbers.

\subsection{Form factor}
The study of form factor is extensively preformed in importance as probe of the structure of a particle as higher energies and wider energy regions become available. As shown in the preceding subsection, the symmetry analysis has deepen our understanding of coupling constants, how to intensify the descriptions of form factor is therefore of particular interest.

The notion of form factor is frequently mentioned in phenomenological description, although it is very difficult to make this notion theoretically precise. Various attempts have been made to improve the consistency between a proposed model and experimental measurements. There is always a plethora of possible choices to construct form factor, herein we resort to the description of radiative transition amplitude, by virtue of which it is possible to isolate some structural features due to quark components of a certain meson, and consequently to factor out the distinctive part for the different final state.

The magnetic dipole transition has been applied to the decay of a vector meson into a pseudoscalar meson and a photon at an early stage of the development of the quark model~\cite{Becchi1965,Becchi1966}. As an approximation~\cite{odonnell1981}, the decay rate can be evaluated in CMS as
\beq
\Gamma (V\to P \gamma) = \frac{2}{3}\alpha |\vec{k}|^3 \left( \frac{E}{m_V} \right) |G(\mu_u, \mu_d, \mu_s)|^2~,
\label{dkrtofvpgma}
\eeq
with
\beq
G(\mu_u, \mu_d, \mu_s)= \sum_{q} \langle P | \hat{M}_q | V \rangle ~.
\label{defofampvp}
\eeq
Here $E=(m^2_V+m^2_P)/2m_V$ and $|\vec{k}|=(m^2_V-m^2_P)/2m_V$; $\hat{M}_q$ is the magnetic momentum operator for the quark $q$ and defined as
\beq
\hat{M}_q \equiv (\hat{Q}_q \mu_q) \cdot \hat{\sigma}_q ~,
\label{defofmq}
\eeq
where $\hat{Q}_q$ and $\hat{\sigma}_q $ are respectively the charge and spin operators whose eigenvalue equations read
\beq
\left.\begin{array}{rl}
\hat{Q}_{q^{\prime}} | q \rangle = & \delta_{q^{\prime} q} e_q | q \rangle ~, \\
\hat{\sigma}_q | \uparrow \rangle = & | \uparrow \rangle ~,~  \hat{\sigma}_q | \downarrow \rangle = - | \downarrow \rangle ~.
\end{array}\right.~~
\label{egneqforqandsgm}
\eeq
Here $e_q $ is the quark charge in unit of the electron charge $e$, and $| \uparrow \rangle$ and $|\uparrow \rangle$ are two independent spin states of a quark.  Notice that if $\bar{q}$ denotes the anti-quark of $q$, $e_{\bar{q}} =- e_q $, which indicates $\hat{Q}_{\bar{q}} =-\hat{Q}_q$. The magnetic momentum of a quark is defined as~\cite{odonnell1981}
\beq
\mu_q \equiv \frac{e \hbar}{2 m_q c}~.
\label{defofqmm}
\eeq

\begin{table}[hbt]
\caption{\label{quarkrptn}The representations of quark-antiquark combinations for $u$, $d$, and $s$ quarks.} \center
\begin{tabular}{ll}\hline \hline \\
$\rop    $,~$ \pip$   & $-u\bar{d}$   \\
$\roz    $,~$ \piz$   & ${\displaystyle \frac{1}{\sqrt{2}}(u\bar{u}-d\bar{d}) }$    \\
$\rom    $,~$ \pim$   & $d\bar{u}$    \\
$\kstp   $,~$ \kap$   & $u\bar{s}$    \\
$\kstz   $,~$ \kaz$   & $d\bar{s}$    \\
$\kstzb  $,~$ \kazb$  & $-s\bar{d}$   \\
$\kstm   $,~$ \kam$   & $s\bar{u}$   \\
$\omega_8$,~$ \eta_8$ & ${\displaystyle \frac{1}{\sqrt{6}}(u\bar{u}+d\bar{d}-s\bar{s}) }$     \\
$\omega_1$,~$ \eta_1$ & ${\displaystyle \frac{1}{\sqrt{3}}(u\bar{u}+d\bar{d}+s\bar{s}) }$    \\
$\omega  $,~$ \etap$  & ${\displaystyle \sin\alpha(-s\bar{s})+\cos\alpha
\left(\frac{u\bar{u}+d\bar{d}}{\sqrt{2}}\right) }$   \\
$\phi    $,~$ \eta$   & ${\displaystyle \cos\alpha (-s\bar{s})-\sin\alpha \left(\frac{u\bar{u}+d\bar{d}}{\sqrt{2}}\right) }$  \\  \hline \hline
\end{tabular}
\end{table}

Exhibited in Table~\ref{quarkrptn} are $SU(3)$ representations appropriate to quark-antiquark combinations forming s-state mesons. The spin-one and spin-zero mesons are then distinguished by their spin wave functions. As examples, the wave functions for $\rop$ and $\pip$ read
\beq
\left.\begin{array}{rl}
 | \rop \rangle = & {\displaystyle (-u\bar{d})
 \cdot \frac{1}{\sqrt{2}} ( |\uparrow\rangle |\downarrow\rangle + |\downarrow\rangle |\uparrow\rangle ) } ~, \\
 | \pip \rangle = & {\displaystyle (-u\bar{d})
 \cdot \frac{1}{\sqrt{2}} ( |\uparrow\rangle |\downarrow\rangle - |\downarrow\rangle |\uparrow\rangle ) } ~. \end{array}\right.~~
\label{wafnofroandpi}
\eeq
With these expressions, the decay amplitude $A_{V \gamma P }$ can be evaluated as follows
\beq
\left.\begin{array}{l}
{\displaystyle \sum_{q} \hat{M}_q | \rop \rangle }= {\displaystyle e_u \mu_u (-u\bar{d})
 \cdot \frac{1}{\sqrt{2}} ( |\uparrow\rangle |\downarrow\rangle - |\downarrow\rangle |\uparrow\rangle ) } ~\\
~~~~~~~~~~~~~~ + {\displaystyle e_{\bar{d}} \mu_d (-u\bar{d})
 \cdot \frac{1}{\sqrt{2}} ( -|\uparrow\rangle |\downarrow\rangle + |\downarrow\rangle |\uparrow\rangle ) } ~\\
 = {\displaystyle (e_u \mu_u + e_d \mu_d ) (-u\bar{d})
 \cdot \frac{1}{\sqrt{2}} ( |\uparrow\rangle |\downarrow\rangle  - |\downarrow\rangle |\uparrow\rangle ) } ~,
\end{array}\right.~~
\label{calculation}
\eeq
therefore
\beq
A_{\pip \gamma \rop } = \sum_{q} \langle \pip | \hat{M}_q | \rop \rangle = \frac{1}{3}(2\mu_u-\mu_d)~.
\label{defofampvp}
\eeq
The other amplitudes can be obtained analogously~\cite{odonnell1981}, and all relevant results are summarized in Table~\ref{dkmtxofvp}.

\begin{table}[hbt]
\caption{\label{dkmtxofvp}Relative strengths of Amplitude elements for magnetic dipole transitions~\cite{odonnell1981}. $s_M \equiv \sin \alpha_M$, $c_M \equiv \cos \alpha_M$, with $M=V,P$.
Here the mixing angle is defined as $\alpha_M \equiv \theta_M - \hat{\theta}$, with $\hat{\theta}=\arctan (1/\sqrt{2}) $. It should be noted that $s_M$ and $c_M$ in this table have the distinctive meanings from $s_{\alpha}$ and $c_{\alpha}$ in Table~\ref{vpmsnfmmix}, although their appearance is identical.}
\center
\begin{tabular}{ll}\hline \hline
Decay mode    & Matrix element          \\ \hline
$\rho^{\pm,0} \to \pi^{\pm,0} \gamma$
              & ${\displaystyle \frac{1}{3}(2\mu_u-\mu_d) }$    \\
$(\kstzb, \kstz) \to (\kazb,\kaz) \gamma$
              &${\displaystyle -\frac{1}{3}(\mu_s+\mu_d) }$    \\
$K^{*\pm} \to K^{\pm}\gamma$
              & ${\displaystyle -\frac{1}{3}(\mu_s-2\mu_d) }$     \\
$\phi \to \eta \gamma$
 &${\displaystyle -\frac{2}{3} \mu_s c_V c_P +\frac{1}{3}(2\mu_u-\mu_d) s_V s_P }$  \\
$\phi \to \etap \gamma$
 &${\displaystyle -\frac{2}{3} \mu_s c_V s_P -\frac{1}{3}(2\mu_u-\mu_d) s_V c_P }$  \\
$\omega \to \eta \gamma$
 &${\displaystyle -\frac{2}{3} \mu_s s_V c_P -\frac{1}{3}(2\mu_u-\mu_d) c_V s_P }$  \\
$\etap \to \omega \gamma$
 &${\displaystyle -\frac{2}{3} \mu_s s_V s_P +\frac{1}{3}(2\mu_u-\mu_d) c_V c_P }$  \\
$\roz \to \eta \gamma$
              & ${\displaystyle -\frac{1}{3}(2\mu_u+\mu_d) s_P }$     \\
$\etap \to \roz \gamma$
              & ${\displaystyle \frac{1}{3}(2\mu_u+\mu_d) c_P }$      \\
$\phi \to \piz \gamma$
              & ${\displaystyle -\frac{1}{3}(2\mu_u+\mu_d) s_V }$     \\
$\omega \to \piz \gamma$
              & ${\displaystyle \frac{1}{3}(2\mu_u+\mu_d) c_V }$      \\
\hline \hline
\end{tabular}
\end{table}

\begin{figure}[htb]
\begin{minipage}{8cm}
\includegraphics[width=3.25cm,height=3.5cm]{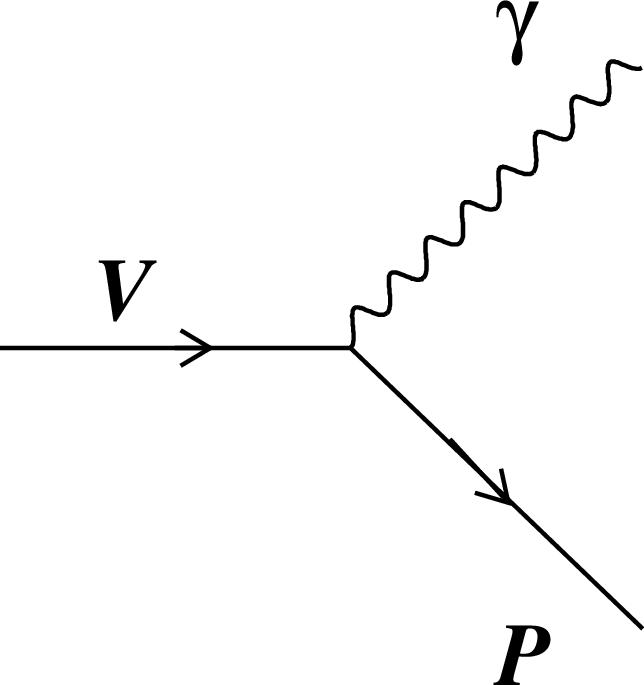}
\hskip 0.5cm
\includegraphics[width=3.25cm,height=3.5cm]{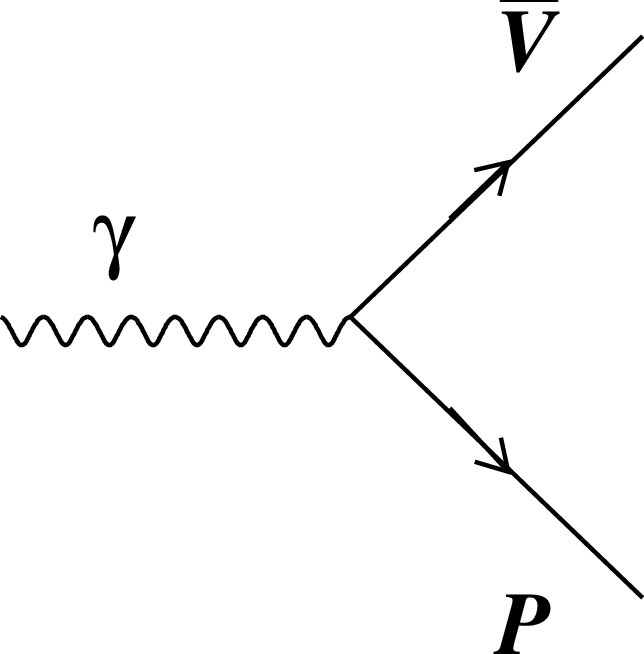}
\end{minipage}
\caption{\label{twoymn} Two amplitudes: one is the decay amplitude $A_{V \gamma P }$ (a), and the other is the production amplitude $A_{\gamma V P } $ (b) from the $\EE$ colliding experiment.}
\end{figure}

In order to find the relation between two kinds of amplitudes, we write them in the current form. One is the decay amplitude $A_{V \gamma P } $,
\beq
A_{V\gamma P}=\langle P| J_{\mu} | V\rangle \propto f_{\gamma P}(s) ~,
\label{ampsvpgma}
\eeq
the other is the production amplitude $A_{\gamma V P } $ from the $\EE$ colliding experiment,
\beq
A_{\gamma V P}=\langle\overline{V} P | J_{\mu} | 0 \rangle \propto f_{V P}(s) ~,
\label{ampsgmavp}
\eeq
where $p$ indicates the momentum of the meson, and $\overline{V}$ indicates the antiparticle of the particle $V$. According to the crossing symmetry~\cite{Feynman1972}, the correspondence between $A_{V \gamma P }$ and $A_{\gamma V P }$ is reciprocal, referring to Fig.~\ref{twoymn} for their Feynman diagrams.

If assume
\beq
\frac{ A_{\gamma V P } }{A_{V \gamma P }  }= C(s) ~,
\label{ratiooftwoamps}
\eeq
thus it is obtained immediately
\beq
\left.\begin{array}{rcl}
f_{V P}(s) &=&  f_{\gamma P}(s)  \cdot C(s) ~\\
           &=&  g(s)  \cdot G(\mu_u, \mu_d, \mu_s)  ~,
\end{array}\right.~~
\label{rlnoftwofmfts}
\eeq
where $G(\mu_u, \mu_d, \mu_s)$ is the function of quark magnetic momentum (given in the left column of Table~\ref{dkmtxofvp}), which is manifestation of structural feature of the constituent quarks. The function $g(s)$ synthesizes all energy ($s$) dependent effects in both $C(s)$ and $f_{\gamma P}(s) $.

\section{Summary}\label{xct_sum}
Our objective is to present the cross section and parametrization for charmonium mesonic decay on an $\EE$ collider, so as to aid the systemically analysis and understand of more and more accumulated charmonium data at the BEPCII and BESII.

Unlike many previous works that are noncommittal on the subject of the magnitude of the cross section, not only angular distributions but also the accurate differential cross sections are calculated in this paper. Moreover, the symmetry analysis and magnetic transition description are utilized and incorporated to provide a fairly meticulous parametrization forms for the future data analysis of charmonium quasi two body decays. Such kinds of phenomenological efforts will prominently benefit our comprehension of the decay mechanism from many aspects, including the meson mixing angle, form factor, and $SU(3)$ symmetry breading effect. We heartily hope that the present work can stimulate and intensify the research for a deeper insight of the fundamental dynamics of charmonium decay.

Last but not least, although the description of parametrization in this paper is concentrated upon the VP mode, the idea itself has the generic meaning and can be extended  easily to other cases.

\section*{Acknowledgment}
This work is supported in part by National Key Research and Development Program of China under Contracts No.~2023YFA1606003 and No.~2023YFA1606000.

\appendix
\begin{appendix}
\section{}
Here only the Feynman rules relevant to the calculation of section II are collected.

External lines radiates from vertices and receive the following factors~:
\begin{enumerate}
\item For spin-1/2 fermion of momentum $p$ and spin state $s$
    \begin{itemize}
    \item in initial state: $u(p,s)$ on the right~,
    \item in final state: $\bar{u}(p,s)$ on the left~;
    \end{itemize}
\item For spin-1/2 antifermion of momentum $p$ and spin state $s$
    \begin{itemize}
    \item in initial state: $\bar{v}(p,s)$ on the right~,
    \item in final state: $v(p,s)$ on the left~;
    \end{itemize}
\item For spin-0 boson
    \begin{itemize}
    \item in either initial or final state: 1~;
    \end{itemize}
\item For spin-1 boson of helicity $\lambda$ (if massless boson, $\lambda=\pm 1$; if massive boson, $\lambda=0, \pm 1$)
    \begin{itemize}
    \item in initial state: $\epsilon_{\mu}(\lambda)$~,
    \item in final state: $\epsilon^*_{\mu}(\lambda)$~.
    \end{itemize}
\end{enumerate}

Internal lines represent the propagation of virtual particles between vertices. Each is associated with a propagator which depends on the momentum of the particle and is diagonal in internal labels~:
\begin{enumerate}
\item For photon or gluon
\beq
i \Delta_{\mu\nu}(p)= \frac{i}{p^2+i \epsilon} \left[ -g_{\mu\nu}+(1-\xi)\frac{p_\mu p_\nu}{p^2} \right]~;
\label{pgtofpton}
\eeq
\item For massive vector boson
\beq
i \Delta_{\mu\nu}(p)= \frac{i}{p^2-m^2+i \epsilon}
\left[ -g_{\mu\nu}+\frac{(1-\xi)p_\mu p_\nu}{p^2-\xi m^2} \right]~.
\label{pgtofmvt}
\eeq
\end{enumerate}

The metric tensor reads
\beq
g_{\mu\nu}=
\left(\begin{array}{cccc}
 1   &  0  &  0 &  0  \\
 0   & -1  &  0 &  0  \\
 0   &  0  & -1 &  0  \\
 0   &  0  &  0 & -1  \\
\end{array}\right) = g^{\mu\nu}~.
\label{mtkgmn}
\eeq

The invariant product of two four-vectors, $p^{\mu}=(p^0,\vec{p})$ and $k^{\mu}=(k^0,\vec{k})$, is defined as
\beq
p_{\mu} k^{\mu} = g_{\mu\nu} p^{\nu} k^{\mu} = p^0 k^0 - \vec{p} \cdot \vec{k}~;
\label{defnofprdt}
\eeq
also the Feynman  ``slash'':
\beq
\rlap/A \equiv  \gamma_{\mu} A^{\mu} = \gamma^{0} A^{0} - \vec{\gamma} \cdot \vec{A}~.
\label{defofslasha}
\eeq

Totally antisymmetric Levi-Civita tensor
\beq
\epsilon_{\mu\nu\lambda\sigma} =
\left\{ \begin{array}{l}
 +1 ~~~\mbox{if \{ $\mu, \nu, \lambda, \sigma $  \} is an even permutation }\\
 ~~~~~~~\mbox{of \{ $0, 1, 2, 3$ \} }   \\
 -1 ~~~\mbox{if it is an odd permutation }  \\
 ~~0  ~~~\mbox{otherwise } \\
\end{array} \right.
\label{defnepsilon}
\eeq

Useful identities
\beq
\epsilon_{\mu\nu\lambda\sigma} \epsilon_{\mu^{\prime}\nu^{\prime}\lambda^{\prime}\sigma^{\prime}}= -
\left|\begin{array}{cccc}
g_{\mu\mu^{\prime}}    & g_{\mu\nu^{\prime}}    & g_{\mu\lambda^{\prime}}    &g_{\mu\sigma^{\prime}}     \\
g_{\nu\mu^{\prime}}    & g_{\nu\nu^{\prime}}    & g_{\nu\lambda^{\prime}}    &g_{\nu\sigma^{\prime}}        \\
g_{\lambda\mu^{\prime}}& g_{\lambda\nu^{\prime}}& g_{\lambda\lambda^{\prime}}&g_{\lambda\sigma^{\prime}}  \\
g_{\sigma\mu^{\prime}} & g_{\sigma\nu^{\prime}} & g_{\sigma\lambda^{\prime}} &g_{\sigma\sigma^{\prime}}
\end{array}\right|~
\label{twoepsilona}
\eeq

\beq
\epsilon_{\mu\nu\lambda\sigma} \epsilon_{\mu \nu^{\prime}\lambda^{\prime}\sigma^{\prime}}= -
\left|\begin{array}{ccc}
 g_{\nu\nu^{\prime}}    & g_{\nu\lambda^{\prime}}    &g_{\nu\sigma^{\prime}}        \\
 g_{\lambda\nu^{\prime}}& g_{\lambda\lambda^{\prime}}&g_{\lambda\sigma^{\prime}}  \\
 g_{\sigma\nu^{\prime}} & g_{\sigma\lambda^{\prime}} &g_{\sigma\sigma^{\prime}}
\end{array}\right|~
\label{twoepsilonb}
\eeq

\beq
\begin{array}{rcl}
\epsilon_{\mu\nu\lambda\sigma} \epsilon_{\mu \nu\lambda^{\prime}\sigma^{\prime}}
&=&-2~(g_{\lambda\lambda^{\prime}} g_{\sigma\sigma^{\prime}}
-g_{\lambda\sigma^{\prime}} g_{\sigma\lambda^{\prime}}) \\
\epsilon_{\mu\nu\lambda\sigma} \epsilon_{\mu \nu\lambda\sigma^{\prime}}
&=&-6~ g_{\sigma\sigma^{\prime}} \\
\epsilon_{\mu\nu\lambda\sigma} \epsilon_{\mu \nu\lambda\sigma}
 &=&- 24~ !
\end{array}~
\label{twoepsilonc}
\eeq

So far as the calculation involving $\gamma$-matrix is concerned, the details can be referred to Ref.~\cite{quangpham}.

\end{appendix}


\begin{thebibliography}{99}
\bibitem{bes}M.~Ablikim {\em et al.}, (BESIII Collaboration), Nucl. Instr. Meth. A {\bf 614}: 345 (2010).
\bibitem{yellow}CHAO Kuang-Ta, WANG Yi-Fang, Internation Journal of Modern Physics A (Suppl. Issue 1),{\bf 24}: 1 (2009)

\bibitem{Kowalski:1976mc}H.~Kowalski and T.~F.~Walsh,
Phys.\ Rev.\  D {\bf14}, 852 (1976).
\bibitem{Clavelli:1983}L.~J.~Clavelli and G.~W.~Intemann,
Phys.\ Rev.\  D {\bf 28}, 2767 (1983).
\bibitem{Haber}H.~E.~Haber and J.~Perrier, \Journal\PRD{32}{2961}{1985}.
\bibitem{Seiden88}A.~Seiden, H.~F.-W.~Sadrozinski, and H.~E.~Haber, \Journal\PRD{38}{824}{1988}.
\bibitem{nMorisita91}N.~Morisita, I.~Kitamura and T.~Teshima,
Phys.\ Rev.\  D {\bf 44}, 175 (1991).
\bibitem{rBaldini98}R.~Baldini {\em et al.}, Phys. Lett. B {\bf 444}, 111 (1998).
\bibitem{zmy2015}K.~Zhu, X.~H.~Mo, C.~Z.~Yuan, Int. J. Mod. Phys. {\bf A30} (2015) 1550148
\bibitem{Baldini19}R.~B.~Ferroli {\it et al.}, Phys.\ Lett.\  B {\bf 799}, 135041 (2019).
\bibitem{moxh2022}X.~H.~Mo and J.~Y.~Zhang, Phys.\ Lett.\  B {\bf 826}, 136927 (2022).
\bibitem{moxh2023}X.~H.~Mo, P.~Wang and J.~Y.~Zhang, Phys.\ Rev.\  D {\bf 107}, 094009 (2023).
\bibitem{moxh2024}X.~H.~Mo, Phys.\ Rev.\  D {\bf 109}, 036036 (2024).
\bibitem{moxh2025}X.~H.~Mo, Phys.\ Lett.\  B {\bf 861}, 139287 (2025).

\bibitem{chungsu2008}S.~U.~Chung and J.~M.~Friedrich, Phys. Rev. D {\bf 78}, 074027 (2008).
\bibitem{chungsu1998}S.~U.~Chung, Phys. Rev. D {\bf 57}, 431 (1998).
\bibitem{chungsu1993}S.~U.~Chung, Phys. Rev. D {\bf 48}, 1225 (1993); {\bf 56}, 4419 (1997).
\bibitem{chungsu1971}S.~U.~Chung, CERN Yellow Report No. 71-8, 1971.

\bibitem{filippini1995}V.~Filippni, A.~Fontana, and A.~Rotondi, Phys. Rev. D {\bf 51}, 2247 (1998).
\bibitem{zou2003}B.~S.~Zou and D.~V.~Bugg, Eur. Phys. J. A {\bf 16}, 537 (2003).

\bibitem{tosa1976}Y.~Tosa, DPNU-34 (July, 1976).

\bibitem{quangpham}Quang Ho-Kim, Xuan-Yem Pham, {\em Elementary particles and their interactions} (Spinger-Verlag 1998)
\bibitem{griffiths}D.~Griffiths, {\em Introduction to elementary particles}, 2nd ed. (WILEY-VCH Verlag GmbH \& Co. KGaA, Weinheim, 2008)

\bibitem{huangsz2003}S.~Z.~Huang {\em et al.}, Eur. Phys. J. C {\bf 26}, 609-623 (2003).
\bibitem{huangsz2005}S.~Z.~Huang {\em et al.}, Eur. Phys. J. C {\bf 42}, 375-389 (2005).

\bibitem{wangp03hepnp}P.~Wang, C.~Z.~Yuan, and X.~H.~Mo, High Energy Phys. Nucl. Phys. 27, 465 (2003).
\bibitem{rudaz}S.~Rudaz, \Journal\PRD{14}{298}{1976}.
\bibitem{wymcgam}P.~Wang, C.~Z.~Yuan, X.~H.~Mo, and D.~H.~Zhang,
\Journal\PLB{593}{89}{2004}.
\bibitem{Wang:2005sk}P.~Wang, X.~H.~Mo, and C.~Z.~Yuan,
Int.\ J.\ Mod.\ Phys.\  A {\bf 21}, 5163 (2006). 
\bibitem{wymphase}P.~Wang, C.~Z.~Yuan, and X.~H.~Mo,
\Journal\PRD{69}{057502}{2004}.
\bibitem{wymogpiapp}P.~Wang, X.~H.~Mo, and C.~Z.~Yuan,
\Journal\PLB{557}{192}{2003}.

\bibitem{pdg2020}P.~A.~Zyla {\em et al.} (PDG), Porg. Theor. Exp. Phys. {\bf 2020}, 083C01 (2020).

\bibitem{Becchi1965}C. Becchi and G. Morpurgo, Phys. Rev. 140, B687 (1965).
\bibitem{Becchi1966}C. Becchi and G. Morpurgo, Phys. Rev. 149, 1284 (1966).
\bibitem{odonnell1981}P.~J.~O'Donnell, Rev. Mod. Phys. {\bf 53}, 673 (1981).

\bibitem{Feynman1972}R.P.~Feynman, {\em Photon-Hadron Interactions} (W.A.~Benjamin Inc., Massachusetts, 1972).


\end{thebibliography}
\end{document}